\documentclass[11pt]{article}
\usepackage{times}
\usepackage{mathptmx}

\usepackage{amsmath,amssymb}
\usepackage{latexsym}

\begin{document}
\title{Non-commutative heat kernel}
\author{Dmitri V. Vassilevich\thanks{Also at V.~A.~Fock Institute of Physics,
St.~Petersburg University, Russia. 
Email: Dmitri.Vassilevich@itp.uni-leipzig.de.
}\\ {\it Institut f\"{u}r Theoretische Physik,
Universit\"{a}t Leipzig,}\\ {\it Augustusplatz 10, D-04109 Leipzig, Germany }
}
\date{\small Received 17 October 2003, revised 6 February 2004}
\maketitle

\begin{abstract}
We consider a natural generalisation of the Laplace type operators
for the case of non-commutative (Groenewold-Moyal star) product. 
We demonstrate existence
of a power law asymptotic expansion for the heat trace of such operators
on $T^n$. First four coefficients of this expansion are calculated
explicitly. We also find an analog of the UV/IR mixing phenomenon when
analysing the localised heat kernel. \end{abstract}


\section{Introduction}
Quantum field theory on non-commutative spaces (see 
reviews \cite{Douglas,Szabo}) is a very fast developing
topic. Although the heat kernel expansion is an essential ingredient
of some earlier approaches to the non-commutative field theory
\cite{Chamseddine:1996zu} properties of the heat expansion of 
operators on non-commutative spaces remain a relatively neglected
subject. We can mention a recent work \cite{Tierz:2003hk} which
studied $q$-deformed zeta functions
(without relation to a particular operator, however). 

At the same time, the heat trace expansion (which is also called
the heat kernel expansion in the physical literature) 
\cite{HKb1,HKb2,Vassilevich:2003xt} is a very powerful instrument
of ordinary (commutative) quantum field theory. 
In particular, coefficients of this expansion define the one-loop
counterterms, quantum anomalies, and various expansions of the
effective action (e.g., the large mass expansion). 

The aim of this paper is to study the asymptotics of the heat trace
for a natural non-commutative generalisation of the Laplace type operator.
Roughly speaking, this generalisation is achieved by replacing ordinary
product by the Groenewold-Moyal 
star (cf. eqs.\ (\ref{starprod}) and (\ref{oper})
below). Because of the presence of the Groenewold-Moyal star the operator
does not fall into
the category considered by Seeley \cite{Seeley67} (see \cite{Grubb02}
for a recent review). Therefore, very little is known about behaviour
of main spectral functions for such operators.  In the next section
we show that somewhat surprisingly the heat trace on a torus
admits a power law
asymptotic expansion for small proper time $t$. The coefficients of
this expansion can be easily calculated. We present explicit expression
for first four coefficients. Our main message is that the heat trace
coefficients for the non-commutative case are fully defined by the
the heat trace expansion for ordinary (``commutative'') but non-abelian
operators.
In section 3 we analyse the localised heat
trace and find an analog of the so-called UV/IR mixing phenomenon. 
\section{Heat trace asymptotics}
Let us consider an $n$-dimensional torus $T^n$ with the coordinates
$0\le x^j <2\pi r_j$, $j=1,\dots,n$. A non-commutative
version of $T^n$ (also denoted as $T^n_\theta$) is constructed
in the following way \cite{Connes:ji}.
One considers a non-commutative associative $*$-algebra with unit
generated by $n$ elements $U_j$ which obey the relation
\begin{equation}
U_j U_l = e^{-i\theta^{jl}/(r_jr_l)} U_l U_j \label{UU}
\end{equation}
with some antisymmetric matrix $\theta^{jl}$. The ``completion'' of
this algebra consists of formal power series with sufficiently fast
decreasing coefficients. There is a one-to-one correspondence between
elements of these algebra and complex valued functions on $T^n$ so that
$U_j$ is identified with $e^{ix^j/r_j}$ and the relation (\ref{UU})
is induced by
the Groenewold-Moyal star product 
\begin{equation}
f\star g = f(x) \exp \left( \frac i2 \, \theta^{\mu\nu}
\overleftarrow{\partial}_\mu \overrightarrow{\partial}_\nu \right)
g(x) \,.\label{starprod}
\end{equation}
In this form the star product
has to be applied to plane waves and then extended
to all (square integrable) functions by means of the Fourier series.
This is exactly the way the star product will be treated in this
paper. We note that (\ref{starprod}) 
arises in the context of deformations of flat Poisson manifolds
\cite{Bayen:1977ha} (see also recent reviews 
\cite{DS02,GPPT}).

In this paper we consider a natural generalisation of the
Laplace type operators for the non-commutative case:
\begin{equation}
D\phi = -\left( \delta^{\mu\nu} \partial_\mu \partial_\nu + a^\mu
\partial_\mu + b \right) \star \phi \,. \label{oper}
\end{equation}
We shall call the operator (\ref{oper}) the star-Laplacian.
We suppose that $\phi$ is multi-component, and $a^\mu$
and $b$ are some matrix valued functions. 
In principle, this construction can be translated into the
vector bundle language.
The coefficient in
front of the second derivative term defines a Riemannian metric
on $T^n$ which is the unit one in the present case. Consequently,
there is no distinction between upper and lower vector indices.  
The operator (\ref{oper}) can be represented in the canonical
form:
\begin{equation}
D\phi =-\left( \delta^{\mu\nu} \nabla_\mu \star \nabla_\nu +
E \right) \star \phi \,,\label{coper}
\end{equation}
where
\begin{equation}
\nabla_\mu =\partial_\mu +\omega_\mu \,,\qquad
\omega_\mu = \frac 12 a_\mu \,,\qquad
E=b-\partial^\mu \omega_\mu -\omega^\mu \star \omega_\mu \,.
\label{omE}
\end{equation}
It is convenient to introduce the field strength of $\omega$:
\begin{equation}
\Omega_{\mu\nu}=\partial_\mu \omega_\nu - \partial_\nu \omega_\nu
+\omega_\mu \star \omega_\nu - \omega_\nu \star \omega_\mu \,.
\label{Omega}
\end{equation}

The inner product in the space of fields is not sensitive to the 
non-commutativity parameter:
\begin{equation}
\langle \psi ,\phi \rangle = \int d^n x \, \psi^\dag \phi
=\int d^n x \, \psi^\dag \star \phi \label{inprod}
\end{equation}
for smooth $\psi$ and $\phi$ (meaning that this star product is closed).
One can easily check that the operator
$D$ is hermitian if $E(x)$ is a hermitian matrix, and $\omega_\mu (x)$
is an anti-hermitian matrix at each point.

Let us now turn to the heat trace
which is defined as a functional
trace in the space of square integrable functions on $M$:
\begin{equation}
K(t,D)={\rm Tr}_{L^2} (\exp (-tD) ) \,. \label{hkern}
\end{equation}
If $D=D_0$ is a partial differential operator of Laplace type (which
is achieved in the limit $\theta \to 0$) the heat kernel is well
defined for positive $t$ and there is an asymptotic series as
$t\to +0$:
\begin{equation}
K(t,D_0)\cong \sum_{k\ge 0} t^{(k-n)/2} a_k(D_0) \,.\label{asymp}
\end{equation}
If $M$ has no boundaries, odd-numbered coefficients vanish,
$a_{2j+1}=0$. Moreover, the coefficients $a_k$ are locally
computable, i.e. they can be presented as integrals over $M$ of
some local invariants constructed from $\nabla_\mu$ and $E$. In
Quantum Field Theory language this corresponds to locality of the
counterterms. We stress, that all these properties hold only in
the limit $\theta\to 0$.

For non-zero values of $\theta$ the operator $D$ is a second order
differential operator on the non-commutative torus. To analyse 
spectral functions of this operator we need a non-commutative
version of the pseudo-differential calculus (which is available
in some form \cite{Connes:ji}), but also a non-commutative
version of the symbolic calculus (which, roughly speaking, tells
how one multiplies symbols of the operators). This second tool
is not available. Therefore, we adopt a different point of view
widely accepted in the quantum field theory context
by considering the star multiplication by a function as being
a pseudodifferential operator on a commutative torus.
The formula (\ref{starprod}) is valid as it stays for the Fourier
harmonics only, but this is exactly what we need in the pseudodifferential
context.
Then $D$ itself is a pseudodifferential operator
($\psi$do) rather than a differential one. The study of spectral
geometry of $\psi$do's was initiated by Seeley \cite{Seeley67}
(see \cite{Grubb02} for an overview). In particular, it was shown
that $\ln t$ terms can appear in the heat kernel expansion and the
heat trace coefficients become, in general, non-local. However,
even these results are not applicable to our case since the symbol
of $D$ does not belong to the so-called standard symbol
space\footnote{Roughly speaking, the symbol $A(x,\xi )$ is obtained from
a $\psi$do $D$ by replacing all partial derivatives by $i\xi$,
like in doing the Fourier transform. For the symbols belonging to
the standard symbol space $\partial^\alpha_\xi A(x,\xi ) =
\mathcal{O} \left( (1+|\xi |^2)^{(m - |\alpha |)/2} \right)$,
where $m$ is the order of $D$ and $\alpha$ is a multi-index. In other words,
each derivative with respect to $\xi$ improves behaviour of $A$ at
$\xi \to \infty$. This condition clearly excludes oscillatory behaviour
of the symbol.}.

To proceed further we need the following definition. We call a
functional of $\nabla_\mu$ and $E$ a star-local polynomial
functional if it is an integral over $M$ of a finite sum of monomials
each consisting of a star product of a finite number of $\nabla_\mu$
and $E$ taken in an arbitrary order. For example, integrals
of $E$ and of $\Omega_{\mu\nu}\star \Omega^{\mu\nu}$ are star local
polynomial functionals, while that of $E^3$ is not. 
In other words, these functionals are are integrals of 
free polynomials of $E$,
the curvature coefficients and their covariant derivatives
(evaluated in the star-algebra).
This definition 
will be useful since we have two different multiplications in the game.

Before analysing the heat trace asymptotics one has to make sure
that the heat trace exists for positive $t$.
For a physicist, the exponential damping established below
is probably enough to establish the existence. There is also a
mathematical proof\footnote{Such a proof was suggested by the referee
of this paper. It consists of the following main steps. 
It is easily seen that the operator $A=D-D_0$ is of degree $1$ in the Sobolev
scale ($AH^s\subset H^{s-1}$). This also ensures existence of the
semigroup $e^{-tD}$. Then if $(\partial_t-D)f=0$, $f\in H^{r,s}$ 
($H^{r,s}$ consists of functions with $r$ square integrable derivatives
w.r.t. $t$ and $s$ w.r.t. $x$), we have $Af, D_0f \in H^{r,s-1}$,
so $f\in H^{r+1,s+1}$.
Thus  for all $t>0$, $e^{-tD}$ is a continuous map 
$L^2\to C^\infty$ so it is
of trace class.\label{footnote}
}. 

Now we can formulate our main result. Let $D$ be a star-Laplacian
(\ref{coper}) on $T^n$. Then
\begin{enumerate}
\item There is a power-law asymptotic expansion (\ref{asymp}) of
the heat kernel for the operator $D$.
The coefficients $a_k(D)$ are star-local polynomial functionals\footnote{
This also implies that all dependence on 
$\theta$ is hidden in the star product. Such property cannot be
supported in the approaches which consider perturbative expansions
in $\theta$.} of $\nabla$ and $E$.
\item In particular,
\begin{eqnarray}
&&a_0= (4\pi)^{-n/2}\, {\rm tr} (I)\, \mbox{volume}\, T^n\,, \label{a0} \\
&&a_2= (4\pi)^{-n/2} \int d^n x \,{\rm tr} (E) \,,\label{a2} \\
&&a_4= (4\pi)^{-n/2} \frac 1{12} \int d^n x \,{\rm tr} \left( 6
E\star E + \Omega^{\mu\nu} \star \Omega_{\mu\nu} \right)
\label{a4}\\
&&a_6= (4\pi)^{-n/2} \frac 1{360} \int d^n x \,{\rm tr} \left(
60 E\star E \star E + 30 E \star E_{;\mu\mu}  \right. \nonumber\\
&&\qquad\qquad\qquad 
+30 E \star \Omega_{\mu\nu} \star \Omega_{\mu\nu} 
- 4 \Omega_{\mu\nu;\rho} \star \Omega_{\mu\nu;\rho}
+ 2 \Omega_{\mu\nu;\nu} \star \Omega_{\mu\rho;\rho}
\nonumber \\
&&\qquad\qquad\qquad \left.
-12 \Omega_{\mu\nu}\star \Omega_{\nu\rho} \star \Omega_{\rho\mu}
\right)\,,
\label{a6}
\end{eqnarray}
where semicolon denotes covariant differentiation, 
$E_{;\mu}:=\partial_{\mu} E +\omega_\mu \star E - E \star \omega_\mu$.
${\rm tr}$ is the matrix trace.
\end{enumerate}

Proof consists in a rather straightforward evaluation of the
asymptotic behaviour of (\ref{hkern}) (cf sec.\ 4.1 of ref.\ 
\cite{Vassilevich:2003xt}).
To calculate the trace in (\ref{hkern}) we need a basis in $L^2$ on the
torus.
Let $\{ u^a \}$ be a basis in the ``internal'' space. Then the functions
\begin{equation}
\phi_k^a (x) = \frac {u^a\, e^{ikx}}{(2\pi )^{n/2} (r_1r_2\dots r_n)^{1/2}}  
\label{basis} 
\end{equation}
with $\{ \tilde k_\mu \} = \{ k_\mu r_\mu \} \in \mathbb{Z}^n$ (no summation
over $\mu$)
form an orthonormal system on the torus. 
One can represent the heat trace
in the form:
\begin{equation}
K(t,D) = \int d^n x \sum_a \sum_k \phi_k^{a\dag}(x) \exp (-tD)
\phi_k^a (x) \,.\label{1K}
\end{equation}
Here the integral over $x$ comes from the scalar product (\ref{inprod})
needed to calculate diagonal matrix elements, the trace is then taken
by summing over $a$ and $k$. As a next step, we expand the exponential
in (\ref{1K}) and push all derivatives to the right. Typical monomial
obtained in this way reads:
\begin{equation}
a\star b \star \dots \star c \star (\partial )^{(\alpha)} \,,\label{tyterm}
\end{equation}
where $(\alpha)$ is a multi-index. $a$, $b$, $c$ stay instead of $E$ or 
$\omega$, or their derivatives. Let now the operator (\ref{tyterm}) act
on $e^{ikx}$. One obtains:
\begin{equation}
e^{ikx} (a \star_k (b\star_k \dots \star_k (c \star_k 1))\dots ) 
(ik)^{(\alpha)}\,, \label{ty2}
\end{equation}
where
\begin{equation}
f\star_k g = f(x) \exp \left( \frac i2 \, \theta^{\mu\nu}
\overleftarrow{\partial}_\mu \overrightarrow{(\partial_\nu +ik_\nu)} \right)
g(x) \,. \label{stark}
\end{equation}
Since this binary operation is not associative we have had to put brackets in 
(\ref{ty2}). Again, the formula (\ref{stark}) has to be understood through
the Fourier series.

Let us expand $a$, $b$, $c$ in Fourier series:
$a(x)=\sum_{q^{[a]}} a_{q^{[a]}} e^{i{q^{[a]}}x}$ etc. If the monomial
(\ref{tyterm}) is sandwiched between $\phi_k^\dag (x)$ and $\phi_k(x)$
the exponential $e^{ikx}$ in (\ref{ty2}) is cancelled, and the 
whole $x$-dependence resides in the phase factor:
\begin{equation}
\exp \left( i (q_\mu^{[a]}+q_\mu^{[b]}+\dots +q_\mu^{[c]}) x^\mu \right)\,.
\label{xdep}
\end{equation}
If now we integrate over $x$ as prescribed by (\ref{1K}), we obtain a
delta-symbol:
\begin{equation}
\delta \left( q_\mu^{[a]}+q_\mu^{[b]}+\dots +q_\mu^{[c]}\right)
\label{delq}
\end{equation}
Next we note that the only effect of the modification (\ref{stark})
of the star product in (\ref{ty2}) is the phase factor
\begin{equation}
\exp \left( -\frac i2 \theta^{\mu\nu}(q_\mu^{[a]}+q_\mu^{[b]}+\dots 
+q_\mu^{[c]}) k_\nu \right) =1 \,,\label{zerophase}
\end{equation}
where we have used (\ref{delq}). Therefore, we can as well delete the
subscript $k$ in the star products in (\ref{ty2}). This technical observation
turns out to be very important. 

Next we collect again all monomials to an exponent. Up to this point we only
used the definition of the exponential in terms of the power series 
forth and back and operated with finite sums or individual monomials.
Let us perform the 
summation over $a$ in (\ref{1K}). This yields:
\begin{equation}
K(t,D)=(2\pi )^{-n} \int \frac{d^nx}{r_1r_2\dots r_n}\, 
\sum_{\tilde k\in \mathbb{Z}^n}
{\rm tr}\, \exp \left[ t\left( (\nabla_\mu +ik_\mu)\star (\nabla_\mu +ik_\mu )
+E\right) \star \right]\,.
\label{2K} \end{equation}
We stress again that the star product in (\ref{2K}) does not depend on $k$.
Therefore, the asymptotic behaviour of (\ref{2K}) can be evaluated rather
straightforwardly. One has to isolate $e^{-tk^2}$ and expand the rest
of the exponential in a power series. To justify this step one has to use
the assumption above. For a physicists, the presence of the damping term
$-tk^2$ in the exponential is already enough to state that the sum
over $k$ is convergent for positive $t$
and that the subsequent integral over $x$ also
exists at least for some ``good'' $E$ and $\omega$. 
Then one sums over $k$ by
using the formulae:
\begin{eqnarray}
&&\sum_{\tilde k\in \mathbb{Z}^n} e^{-tk^2} 
\cong \left( \frac {\pi}t \right)^{n/2} (r_1 r_2 \dots r_n)
\,,\nonumber \\
&&\sum_{\tilde k\in \mathbb{Z}^n} k_\mu k_\nu e^{-tk^2} 
\cong \delta_{\mu\nu} \frac 1{2t}
\left( \frac {\pi}t \right)^{n/2} (r_1 r_2 \dots r_n)  \,, \label{sumk}\\
&&\sum_{\tilde k\in \mathbb{Z}^n} k_\mu k_\nu k_\rho k_\sigma e^{-tk^2} 
\cong (\delta_{\mu\nu} \delta_{\rho\sigma} + \delta_{\mu\rho}\delta_{\nu\sigma}
+\delta_{\mu\sigma}\delta_{\nu\rho} )
\frac 1{4t^2} \left( \frac {\pi}t \right)^{n/2} (r_1 r_2 \dots r_n)  \,,
\nonumber 
\end{eqnarray}
etc. Corrections to (\ref{sumk}) are exponentially small if $t\to +0$.
Clearly, in this way one obtains a power-law asymptotic expansion
(\ref{asymp}). If, as usual, one assigns dimension one to $\nabla$ and
dimension two to $E$, simple power counting arguments show that
each coefficient $a_p$ is a star-local polynomial functional of $E$ and
$\nabla$ with an integrand of the dimension $p$. 
This proves the first
assertion formulated above in this section.

To illustrate how the procedure works, consider a simplified case $E=0$
and calculate the heat trace coefficients up to $a_4$.
\begin{equation}
K(t,D)=(2\pi )^{-n} \int \frac{d^nx}{r_1r_2\dots r_n}\, 
\sum_{\tilde k\in \mathbb{Z}^n}
{\rm tr}\, e^{-tk^2} 
\exp \left[ t\left( \nabla_\mu \star \nabla_\mu +2ik_\mu \nabla^\mu 
\right) \star \right]\,.
\label{3K} \end{equation} 
Now we expand the second exponential keeping only the terms of
dimension four or lower:
\begin{eqnarray}
&&K(t,D)\cong (2\pi )^{-n} \int \frac{d^nx}{r_1r_2\dots r_n}\, 
\sum_{\tilde k\in \mathbb{Z}^n}
{\rm tr}\, e^{-tk^2} \left[ 1 + t \nabla^\mu \star \nabla_\mu 
\right. \nonumber\\
&&\qquad\qquad -2 t^2 (k\nabla )\star (k\nabla ) +\frac{t^2}2 
\nabla^\mu \star \nabla_\mu \star  \nabla^\nu \star \nabla_\nu
\nonumber \\
&&\qquad\qquad  
-\frac {2t^3}3 \left( (k\nabla )\star (k\nabla )\star 
\nabla^\mu \star \nabla_\mu 
+ \nabla^\mu \star \nabla_\mu \star (k\nabla ) \star (k\nabla)
\right. \\
&&\qquad\qquad \left.\left.
+ (k\nabla ) \star \nabla^\mu \star \nabla_\mu \star (k\nabla)
\right) + \frac{2t^4}3 (k\nabla )\star (k\nabla )\star
(k\nabla )\star (k\nabla ) \right], \nonumber
\end{eqnarray}
where $(k\nabla ):=k^\mu \nabla_\mu$.
Next we use (\ref{sumk}) to perform summation over $k$. All covariant
derivatives combine into commutators yielding the final result:
\begin{equation}
K(t,D)\cong (4\pi t )^{-n/2} \int d^n x \, {\rm tr}
\left[ 1 + \frac{t^2}{12} \Omega_{\mu\nu} \star \Omega^{\mu\nu}
+\mathcal{O}(t^3) \right] \,.\label{4K}
\end{equation}
This result confirms (\ref{a0}) - (\ref{a4}) in the particular case
considered. In principle one can go on and compute the rest of
(\ref{a0}) - (\ref{a6}). However, this is not needed. The crucial
fact is that the calculations go exactly the same way as in the
commutative case (cf.\ sec.\ 4.1 of \cite{Vassilevich:2003xt}).
The reason is that even in a theory is commutative, both $\omega$ and
$E$ are matrix-valued, and, therefore, commutativity is not being
used in the course of the calculations. As a consequence, the heat trace
coefficients (\ref{a0}) - (\ref{a6}) can be read off from the commutative
but non-abelian results presented e.g. in \cite{HKb1,HKb2,Vassilevich:2003xt}.
\section{Localised heat trace and UV/IR mixing}
In ordinary commutative case the ``global'' heat trace
(\ref{hkern}) is sometimes replaced by a more general
(localised) expression
\begin{equation}
K(f;t,D_0)={\rm Tr}_{L^2} \left( f \exp (-tD_0) \right)\,, \label{comloc}
\end{equation}
where $f$ is a function. Obviously, $K(t,D_0)=K(1;t,D_0)$. By varying
(\ref{comloc}) with respect to $f(x)$ one obtains matrix elements
of $\exp (-tD_0)_{x,y}$ at coinciding arguments, $x=y$. This modification
proves convenient for technical reasons \cite{HKb1,HKb2}. More important
is that (\ref{comloc}) describes local quantum anomalies (cf. 
\cite{Vassilevich:2003xt} and references therein).

A natural generalisation of (\ref{comloc}) to the non-commutative
case reads:
\begin{equation}
K_\star (f;t,D)={\rm Tr}_{L^2} \left( f \star \exp (-tD) \right) \,.
\label{nloc}
\end{equation}
Let us consider the $t\to +0$ asymptotic expansion of this expression.
Clearly, all methods used in the previous section work also for
this case. The only modification is to replace $a$ by $f$ in (\ref{tyterm})
and (\ref{ty2}). Therefore, we conclude that there is an asymptotic
expansion
\begin{equation}
K_\star (f;t,D)\cong \sum_{k=0,2,4,\dots} t^{(k-n)/2}
a_k (f,D) \,, \label{smas}
\end{equation}
where the coefficients $a_k(f,D)$ are star-local polynomial functionals.
Again, commutative non-abelian
heat trace coefficients define uniquely the heat trace expansion
for the non-commutative heat trace. One only has to remember to take
matrix-valued smearing function $f$ in the commutative case in order
to preserve all relevant invariants\footnote{One can
calculate the heat trace expansion 
 even if $f$ is a differential operator \cite{Branson:1997ze}.}.
In particular, by comparing with non-abelian commutative heat kernel
coefficients given in \cite{HKb1,HKb2,Branson:1997ze} one easily derives
first three coefficients in the expansion (\ref{smas}):
\begin{eqnarray}
&&a_0(f,D)= (4\pi)^{-n/2}\, {\rm tr} (f)\, \mbox{volume}\, T^n\,, \nonumber \\
&&a_2(f,D)= (4\pi)^{-n/2} \int d^n x \,{\rm tr} (f\star E) \,,
\label{a024} \\
&&a_4(f,D)= (4\pi)^{-n/2} \frac 1{12} \int d^n x \,{\rm tr}
\left[ f\star \left( 
2 E_{;\mu\mu} 
+6E\star E + \Omega^{\mu\nu} \star \Omega_{\mu\nu} \right)\right]
\nonumber 
\end{eqnarray}
Of course, the same expressions may be obtained by more direct
methods (cf. equations (\ref{3K}) - (\ref{4K}) above). One can make
a simple but important observation that the ``zero-momentum'' limit
$f\to 1$ commutes with the asymptotic expansion in $t$ so that
$a_k(D)=a_k(1,D)$.

One can also consider a different generalisation of (\ref{comloc}) to
the non-commutative case:
\begin{equation}
K(f;t,D)={\rm Tr}_{L^2} \left( f [\exp (-tD) ] \right)\,, \label{nc-loc}
\end{equation}
where there is no star between $f$ and the exponent. 

To evaluate an asymptotic expansion of (\ref{nc-loc}) one can use
again the same plane wave basis as in (\ref{1K}), but there is one
important difference. Now we cannot replace the modified product
$\star_k$ by $\star$ under the trace. The reason is that the momentum
corresponding to $f$ enters, of course, the momentum conservation
delta function (cf. (\ref{delq})) but does not appear in an analog
of the phase factor (\ref{zerophase}). Therefore, one cannot
guarantee existence of the power-law asymptotics and the star-locality.
As we will see in a moment,
all these properties are indeed violated. 

To illustrate this point let us calculate (\ref{nc-loc}) in the case
of zero connection $\omega_\mu =0$ to the linear order in $E$ neglecting
also all derivatives of $E$ which are not coupled directly to $\theta$. In
this approximation
\begin{equation}
K(f;t,D)= \int d^n x \sum_a \sum_k \phi_k^{a\dag}(x) f(x) \exp (-tk^2)
\left( t\, E \star \phi_k^a (x)\right) \,.\label{linE}
\end{equation}
After expanding $f$ and $E$ in Fourier series, summing over $k$,
and returning back to the coordinate representation one obtains: 
\begin{equation}
K(f;t,D)= \frac t{(4\pi t)^{n/2}} \int d^nx {\rm tr} \left[ E(x)
\exp \left( \frac{\theta^{\mu\nu} \theta^{\mu \rho} \partial_\nu
\partial_{\rho}}{16t} \right) f(x)\right] +\dots \label{uvir}
\end{equation}
where dots denote the higher order terms which we dropped in this calculation.

The differential operator in the exponential in (\ref{uvir}) is non-positive.
Therefore, for a non-constant $f$ there is a very strong exponential damping
at $t\to 0$. At first glance this exponential looks as a regulator of
the heat kernel. However, the whole effect disappears for $f=const$.
This is a manifestation of the so-called UV/IR mixing \cite{UVI1,UVI2,UVI3}
which is a characteristic feature of non-commutative field theories.
In the heat-kernel context this mixing is rather a consequence of the
way we have made the localisation. The exponential factor in (\ref{uvir})
is similar to the typical exponent $\exp (-(x-y)^2/4t )$ which appears
in matrix elements of the heat kernel of a ``commutative'' operator $D_0$
between non-coinciding points $x$
and $y$. 
\section{Conclusions}
In this paper we considered a natural generalisation of the Laplace
type operators for the non-commutative case 
(which we call the star-Laplacians).
We have demonstrated that
both global (\ref{hkern}) and localised
(\ref{smas}) heat traces for these operators on $T^n$ admit a power-law
asymptotic expansion for $t\to +0$. The coefficients of these expansions
are star-local functionals of the $\omega$ and $E$. They can be
easily calculated. Expressions for fist several coefficients have been
given explicitly. We have also considered a different way (\ref{nc-loc})
to localise the heat trace and observed an analog of of the UV/IR mixing
phenomenon. Most of the results of this paper may be reformulated for
$\mathbb{R}^n$ by imposing suitable fall-off conditions on the fields
and on the function $f$.

We note that the star product is a natural object in the operator theory
which describes composition of symbols of $\psi$do's
(see \cite{Grubb02,BerezinShubin}).
It has been used in calculations of the effective action in {\it commutative}
field theories \cite{Pletnev:1998yu}

Our results ``confirm'' in a way the spectral action principle 
\cite{Chamseddine:1996zu}: the $\Omega$-term in $a_4$ 
(cf. (\ref{a4})) is just the action for non-commutative Yang-Mills theory.

Our results suggest that the  background field formalism and 
spectral regularization methods (like, e.g., the zeta function regulation)
are efficients tools to study divergences in non-commutative theories.
Implications for renormalization of non-commutative field theories
will be analysed in a separate publication.
\section*{Acknowledgements}
In the course of preparation of this paper I benefited from discussions
and correspondence with many colleagues. I am especially grateful to
I.~Arefeva, M.~Buric, A.~Bytsenko, H.~Grosse, G.~Grubb, Yi Liao,
M.~Schweda and D.~Sternheimer. I thank  the anonymous referee for
helpful critical remarks on the previous version of the manuscript and,
especially, for providing a proof of existance of the heat trace (cf.
footnote \ref{footnote}).
This work has been supported in part 
by the DFG project BO 1112/12-1 and by the Erwin Schr\"{o}dinger
Institute for Mathematical Physics (Vienna).

\end{document}